# Characterizing Interdisciplinarity of Researchers and Research Topics Using Web Search Engines

Hiroki Sayama[1,2*] and Jin Akaishi[1,3]
[1] Collective Dynamics of Complex Systems Research Group, Binghamton University, USA
[2] Departments of Bioengineering & Systems Science and Industrial Engineering, Binghamton University, USA
[3] Faculty of Liberal Studies, Kumamoto National College of Technology, Japan
* Corresponding author

**Abstract**

**Researchers' networks have been subject to active modeling and analysis. Earlier literature mostly focused on citation or co-authorship networks reconstructed from annotated scientific publication databases, which have several limitations. Recently, general-purpose web search engines have also been utilized to collect information about social networks. Here we reconstructed, using web search engines, a network representing the relatedness of researchers to their peers as well as to various research topics. Relatedness between researchers and research topics was characterized by *visibility boost*—increase of a researcher's visibility by focusing on a particular topic. It was observed that researchers who had high visibility boosts by the same research topic tended to be close to each other in their network. We calculated correlations between visibility boosts by research topics and researchers' interdisciplinarity at individual level (diversity of topics related to the researcher) and at social level (his/her centrality in the researchers' network). We found that visibility boosts by certain research topics were positively correlated with researchers' individual-level interdisciplinarity despite their negative correlations with the general popularity of researchers. It was also found that visibility boosts by network-related topics had positive correlations with researchers' social-level interdisciplinarity. Research topics' correlations with researchers' individual- and social-level interdisciplinarities were found to be nearly independent from each other. These findings suggest that the notion of "interdisciplinarity" of a researcher should be understood as a multi-dimensional concept that should be evaluated using multiple assessment means.**

Introduction

The structural and dynamical properties of networks among researchers have been an important research subject for the last several decades (1-15). Price originally proposed the key idea of preferential attachment and the resulting scale-free degree distributions for networks of scientific publications (1), which is now widely applied and utilized in various kinds of scientific fields (16-18). Typical data sources for such studies on networks of scientific communities are domain-specific electronic paper or citation archives, such as arXiv.org for physics, DBLP for computer science, and SSRN for social sciences, from which co-authorship or citation networks can be created. Price's predictions have been confirmed in those data, such as scale-free degree distributions and the network growth over time based on the preferential attachment principles (4,5,8).

While earlier studies mostly focused on citation or collaboration networks within a particular domain, there is a growing body of literature on the characterization and measurement of interdisciplinarity of scientific journals and researchers (19-22). These recent studies used cross-disciplinary citation indexing services, such as ISI Web of Knowledge, and analyzed how multiple disciplines are connected by publications and researchers. Interdisciplinarity has been characterized in several different ways, e.g., how many different disciplines were represented in the references cited in a single paper, how many different disciplines an individual researcher publishes his/her work in, and so on (19,22).

However, the existing citation indexing services have several limitations. One apparent limitation is the lack of flexibility in their disciplinary classification. Established disciplinary classification structures, such as those used in ISI Journal Citation Reports, are based on traditional notions of scientific disciplines, which may not be up-to-date for capturing emerging fields of cutting-edge research where the characterization of interdisciplinarity is most needed. Also, it is commonly assumed in the citation indexing services that each journal belongs to just a few disciplines (mostly just one), which is not necessarily a valid assumption when analyzing properties of highly interdisciplinary publications. Another limitation is that their indexing coverage may not include non-mainstream journals, conference proceedings, and other online archives, which are often more important in particular disciplines (e.g., in computer science and physics). Finally, the data of the citation indexing services are only available on a subscription basis, which would be hard to obtain for researchers whose institutions do not have a subscription to those services.

Given those limitations of the citation indexing services mentioned above, researchers have recently started to utilize more general-purpose web search engines as an alternative data source for researchers' network reconstruction (10,14). For example, Lee et al. (14) defined "Google correlation" (i.e., number of hits obtained by a Google search query for names of two persons) and used it to reconstruct social networks of physicists and politicians. They also evaluated the validity of this data collection method by showing that the reconstructed network by Google searches was indeed correlated with the real social network (14).

There are several significant advantages in this new web search engine-based method over the conventional methods. First, it can exploit a massive amount of information about various forms of association between two researchers (or any pair of keywords) that are collectively produced and maintained by people from all over the world. While web search-based data is less structured and more noisy than data collected from the online archives and citation indexing services, the information created through massive "collective intelligence" is often quite informative and useful (23,24). Second, the use of general-purpose web search engines liberates researchers from the existing disciplinary classification structures, giving them full flexibility in choosing any relevant set of disciplinary keywords to study. Third, many non-indexed sources of information can be included in the analysis, such as conference proceedings and online archives. And finally, the data is open and accessible to everyone, with no subscription required. These advantages are quite suitable and beneficial for studying interdisciplinarity of researchers, but to our knowledge, there is no study reported yet on the use of such web search engines for that purpose.

Here we conducted a preliminary study on the interdisciplinarity of individual researchers and a variety of research topics using a web search engine-based data collection method. We searched the web for information about connections between individual researchers *as well as* researchers and research topics, each quantified by the number of hits obtained through a search query for two keywords (names of two researchers, or a name of a researcher and a research topic). One of the novelties of our work compared to earlier literature is that we developed a unique measurement called *visibility boost*, defined as an increase of a researcher's visibility brought by focusing on a particular research topic. We propose this as a more meaningful way of quantifying the relatedness between the researcher and the research topic than simply using the number of web search hits for those two keywords. We also characterized each individual researcher's interdisciplinarity by measuring the diversity of research topics related to him/her as well as his/her centrality in the researchers' network. The former measurement represents individual-level interdisciplinarity, i.e., how diverse the research topics the researcher is associated with, while the latter captures social-level interdisciplinarity, i.e., how important the researcher is in connecting other researchers.

**Methods**

Web search engine-based data collection methods require a list of keywords to be searched for. In our study, we created the following two separate lists. One is a list of names of 1,000 researchers. This list was compiled by having student volunteers manually collect about 4,000 names from four annual international conference websites for years 2006-2009, and then selecting the top 1,000 significant names based on their numbers of independent web search hits. The four selected conferences were all interdisciplinary ones the authors were already personally familiar with. Although our prior familiarity with the conferences could be a source of potential biases, it was necessary in order for us to be able to manually check and correct mistakes in the raw data collected by student volunteers. It was also our hope that the interdisciplinary nature of these conferences would allow us to create a representation of broader research communities within limited time and labor available.

The other list of keywords is a list of research topics. One could use the traditional categories established in scientometrics literature (e.g., ISI Journal Citation Reports categories) for this purpose. We did not take that option, however, because the relevant research topics discussed in those four subject conferences did not quite fit into the ISI JCR categories, and also because one of our objectives was to demonstrate the flexibility in keyword selection. Therefore, the list of research topics were collected again from the same set of websites by the student volunteers, and then manually edited and compiled by the authors. In so doing, we paid attention to maintaining a good balance among different disciplines. Specifically, we set four major categories and made sure that the numbers of keywords were similar across those categories. As a result, we had 13 words for biological and medical sciences, 9 words for physical sciences, 9 words for engineering and robotics, and 9 words for general terms (40 research topics in total). Because there was an overall emphasis on biological systems among the four conferences we used, there were slightly more words in the first category. The actual list of research topics can be found in Appendix.

In total, we compiled 1,000 (researchers' names) + 40 (research topics) = 1,040 keywords to use. The total number of search queries was (1,040 choose 2) – (40 choose 2) = 539,500. The subtraction of "40 choose 2" was because we did not need to measure the relationships between research topics. We recognize that our specific choices of those researchers' names and research topics may have significantly influenced the results described below, which will be discussed in more detail later.

The overview of the network to be reconstructed using these keywords is illustrated in Fig. 1. The data about the relatedness between researchers and research topics (Fig. 1, left) were used to characterize each researcher's unique research areas as well as his/her individual-level interdisciplinarity. The data about the relatedness among researchers (Fig. 1, right) were used to characterize the researcher's social-level interdisciplinarity. Our primary goal is to illustrate how the proposed method works in identifying possible relationships between these two characterizations.

We implemented a computer program in Java for repeated searches using Google Web Search API (25). Google Web Search API allows one to write a program that can send a search query directly to Google web search engines and then receive search results (with some limitations). Earlier work also used Google for social network data collection (10,14,26-29). In every single search query we conducted in this study, we always included an additional word "research" in order to narrow search results to those related to scientific research (although this was by no means a perfect filtering technique), following a similar technique used by Lee et al. (14). To improve the reliability of search results, each query was searched three times at different times in a day, and their average values were used for analysis (see Appendix for details). We inserted sufficient amount of waiting time between queries in order to avoid overloading the search engine. Therefore the actual data collection took place rather slowly over several weeks in June and July 2010.

In characterizing researchers' relatedness with particular research topics, we had to address the following technical problem: More common words tended to result in more search hits regardless of a researcher's actual research domain. For example, the word "biology" is more commonly used than the word "network", and therefore a search query "John Doe" + "biology" + "research" can produce more search hits than "John Doe" + "network" + "research", even if John Doe's research domain is network science and not biology. This means that one cannot simply use the absolute number of search hits for characterizing unique research areas and interdisciplinarity of individual researchers.

We solved this problem by introducing a new quantity, named *visibility boost (VB)*, defined as

$$VB(r,t) = \frac{h(r,t)}{\sum_i h(i,t)} \bigg/ \frac{\sum_j h(r,j)}{\sum_{i,j} h(i,j)} = \frac{h(r,t)\sum_{i,j} h(i,j)}{\sum_i h(i,t)\sum_j h(r,j)}, \qquad (1)$$

where $r$ is the researcher, $t$ the research topic, and $h(r, t)$ the number of search hits for search query "$r$" (researcher's name) + "$t$" (research topic) + "research". This formula mathematically describes how much change occurs to the visibility of researcher $r$ (i.e., ratio between $r$'s own hits and the total hits over all researchers) by limiting the focus to research topic $t$ (Fig. 2). A visibility boost greater than (or less than) 1 means that researcher $r$ is more (or less) associated with research topic $t$ on the web. A similar link weight normalization method was also proposed by Lee et al. (14), though their method produces values that are influenced significantly by frequencies of two keywords searched for. In contrast, our visibility boost gives a more consistent, intuitive measure of association. Specifically, $VB = 1$ always means a neutral level of association between two keywords, while such a constant reference value for neutrality does not exist in Lee et al.'s method. This property allows one to use visibility boost values comparatively for multiple different topics.

**Results**

Figure 3(a) shows the network of 1,000 researchers reconstructed from the data obtained above (which corresponds to the researchers' network illustrated on the right in Fig. 1). Nodes and undirected links represent researchers and their relatedness, respectively, where the average numbers of search hits were used as link weights. Figure 3(b) shows a complementary cumulative distribution of total link weights of nodes, which does not follow a power-law but still shows a remarkably long tail even with this small data set. Note that the following analysis also used a bipartite network made of connections between 1,000 researchers and 40 research topics (illustrated on the left in Fig. 1), which is not visualized here.

To evaluate the utility of the proposed visibility boost, we counted how many unique researchers would be ranked within top 20%, at least once, according to their visibility boosts by any of the 40 research topics. We then conducted the same counting task using the raw search hit counts instead of visibility boosts. In addition, as a control, we also counted how many unique researchers would be selected, at least once, if 20% of researchers were purely randomly sampled 40 times. The results are summarized in Fig. 4(a). Based on the raw search hit counts (yellow), only less than 60% of the researchers had a chance to be ranked within top 20%. This indicates that relying on raw search hit counts would cause unwanted concentration of analysis on fewer researchers with greater general popularity. In contrast, using the visibility boosts for the same task (red) resulted in nearly every researcher having a chance to be ranked within top 20% for some topic, which is comparable to the random sampling case that showed perfect coverage (blue). This result demonstrates that the proposed visibility boost measure is useful in extracting information about individual researchers' unique specialties, without being dominated by general popularity differences among them.

We found, both visually and statistically, that researchers who had high visibility boosts by the same research topic tended to aggregate in their relatedness network. Figure 4(b) presents statistical evidence supporting this observation, in which average shortest path lengths among the selected 20% researchers were calculated for 40 cases and their smoothed histograms were plotted, under two conditions used in Fig. 4(a): selection by visibility boost (red) and pure random selection (blue). The average shortest path lengths among the top 20% researchers under the former condition (red) were significantly shorter than their random counterparts (blue), implying that researchers strongly associated with a particular topic were indeed located closer to each other, possibly forming a research community on that topic.

We calculated correlations between the visibility boosts by research topics for a researcher and his/her overall popularity and individual-level interdisciplinarity. The popularity was measured by *total topic hits (TTH)*, defined as

$$TTH(r) = \sum_j h(r, j), \tag{2}$$

i.e., how many search hits the researcher *r* had in total across all the research topics. Individual-level interdisciplinarity of a researcher was defined in this study as the diversity of research topics associated with him/her. Following similar metrics used in the literature (21,22), we characterized the individual-level interdisciplinarity by *topic hit entropy (THE)*, defined as

$$THE(r) = -\sum_j \frac{h(r, j)}{TTH(r)} \log \frac{h(r, j)}{TTH(r)}, \tag{3}$$

which is a Shannon entropy applied to the frequency distribution of search hits over all the topics. It is small if the researcher is strongly associated with a small number of research topics but not to others, or large if he/she is associated broadly with many research topics.

Figure 5 shows correlation coefficients between the two measurements introduced above and researchers' visibility boosts by various research topics. The topics are sorted from positive to negative correlations. It is observed in Fig. 5(a) that common words tend to correlate positively with the overall popularity of a researcher, while technical terms tend to correlate negatively. This is not surprising, because popular researchers who frequently appear on news and other online media (i.e., those who have high total topic hits) would tend to be associated more with common words on the web. The word order changes, however, when correlations with topic hit entropy are plotted instead (Fig. 5(b)). There was no correlation found between Figs. 5(a) and 5(b) regarding the word positions in the rankings, which means that each research topic has unique, independent effects on popularity and individual-level interdisciplinarity of a researcher. Of particular interest are the research topics that moved significantly from a negative side in Fig. 5(a) to a positive side in Fig. 5(b), such as "evolution", "biology", "neuron", "cognition", "dynamics", "simulation", and "modeling". This implies that researchers who are strongly associated with these topics tend to be less popular overall but associated with diverse topics at an individual level.

Next, we investigated correlations of a researcher's visibility boosts by research topics with his/her social-level interdisciplinarity, i.e., how "central" he/she is in the researchers' network. We considered three typical centrality measurements: degree, betweenness and closeness (30). Some elaboration was required in measuring degree centrality because links in our network were weighted and the weights might be heterogeneously distributed. We used two approaches in measuring degrees. One was to calculate the Shannon disparity of link weights on a node, introduced by Lee et al. (14), which is given by

$$D(r) = \exp\left(-\sum_j \frac{w(r,j)}{TNH(r)} \log \frac{w(r,j)}{TNH(r)}\right) = \prod_j \frac{w(r,j)}{TNH(r)}^{-\frac{w(r,j)}{TNH(r)}}, \quad (4)$$

where $w(r, j)$ is the number of search hits for search query "$r$" (researcher's name) + "$j$" (another researcher's name) + "research", and $TNH(r)$ the *total name hits* defined as

$$TNH(r) = \sum_k w(r,k). \quad (5)$$

Note that this Shannon disparity $D(r)$ is an exponential of a Shannon entropy of the link weight distribution on node $r$, which is the effective number of links of researcher $r$ if link weights were all equal. To make the terminology more intuitive, we call $D(r)$ an *effective degree* of researcher $r$.

The other approach we took in measuring degree centrality is to calculate *total normalized incoming link weights* of a node, defined as

$$I(r) = \sum_j \frac{w(j,r)}{TNH(j)}. \quad (6)$$

This is a sum over *j* of how much portion of *j*'s link weights comes in to *r*, which characterizes how important node *r* is to other nodes.

For the other two centrality measures (betweenness and closeness), the reciprocals of link weights were used as edge distances. We used Python NetworkX's (31) built-in functions to calculate these centralities. We note that the use of betweenness as a measure of interdisciplinarity was already proposed by Leydesdorff (20), but it was the betweenness of a journal in a citation network while ours is the betweenness of an individual researcher in the researchers' relatedness network.

The results are summarized in Fig. 6. We found that the rankings of the topics "complex network" and "social network" jumped up drastically from Fig. 6(a) to Figs. 6(b), 6(c) and (d) (and a single-word topic "network" also showed similar behavior, but in a slightly different way). This implies that the researchers who are strongly associated with these topics tend to be important to other researchers (Fig. 6(b)) and occupy central positions in the network (Figs. 6(c), 6(d)) *without being associated with a broader range of other researchers (Fig. 6(a)) or topics (Fig. 5(b)).* In other words, researchers who are more strongly associated with network-related topics may have higher social-level interdisciplinarity in their network without having too broad social relatedness.

Moreover, we also found that the correlation strengths of a research topic with researchers' individual-level and social-level interdisciplinarities were nearly independent from each other. Figure 7 shows the distributions of research topics in a two-dimensional correlation coefficient space based on the data in Figs. 5(b), 6(b), 6(c) and 6(d), where research topics are widely scattered with no clearly identifiable tendency. This implies that the notion of "interdisciplinarity" should be understood as a multi-dimensional concept and should be evaluated using multiple assessment means, which is consistent with what has been suggested in the interdisciplinarity research literature (19,21,22).

**Discussion**

In this paper, we illustrated our web search engine-based method to reconstruct a network of relatedness between individual researchers and research topics through preliminary data collection and analysis using a small set of keywords. Our novel contributions include the proposal of visibility boost, a new quantity defined for a pair of a researcher and a research topic, which indicates how the research topic helps increase the researcher's visibility.

Our results showed that visibility boosts by research topics correlated in various ways with other metrics. Most notably, the network-related topics increased their rankings in the order of correlations with researchers' social-level interdisciplinarity even though they were not strongly correlated with the researchers' effective degrees. This finding poses an intriguing future research question about potential causal relationships between topics a researcher works on and his/her position and role in a social context. A straightforward interpretation is that network science is currently a hot topic and therefore network researchers may be referred to more often in public media and other online documents, naturally increasing their centrality in our data set. The opposite explanation is also plausible, though, in that researchers who work at the boundaries of different disciplines may tend to choose networks as part of their research subjects because of their generality and broad applicability to many domains. Yet another, somewhat behavioral, explanation would also be possible, in that those who are aware of properties of complex networks may be able to utilize their knowledge and strategically optimize their positions in a social network. The data used in this study did not contain any causal information and therefore no conclusion can be derived at this point. More systematic studies on temporal changes of researchers' networks will help explain the underlying mechanisms responsible for the patterns observed in this study.

Our work also showed that the individual-level and social-level interdisciplinarities may not necessarily behave similarly in terms of their correlations with specific research topics. This fact suggests that the interdisciplinary nature of a research field should be considered as a multi-dimensional construct ranging over multiple levels, taking into account how many different concepts/disciplines are involved, how important the roles played by the researchers in that field are in connecting different research communities, and so on. This insight may be informative for those who work on formal or informal assessments of academic activities of researchers and research institutions.

We must emphasize that our study is still preliminary and it still has several fundamental limitations. First and foremost, the data collection takes a lot of time in the current form of the proposed method. The number of keyword pairs one has to search for grows quadratically with the number of keywords; each pair must be searched for multiple times in order to improve the reliability of results; and search queries must be sent to web search engines at sufficient intervals in order to avoid interfering with their regular operations. This problem put significant constraints on the scalability of our method in this study. A closer collaboration with web search and other IT industries will likely offer technical solutions to this limitation.

The second limitation is the relatively low reliability of data. It is known that numbers of web search hits are often unreliable because of the lack of incentives for web search providers to give an accurate estimate of search hits (for example, see (32)). Moreover, the search results can contain anything on the web, possibly including wrong, irrelevant, and redundant webpages in search hits. We used an additional keyword "research" in every search query to reduce such risks, but it is still far from optimal. Another critical issue is the possibility of multiple people who share an identical name. We manually checked to make sure there were no such names included in our list of researchers. However, this may not be perfect because of inherent difficulty in identifying/distinguishing researchers only by their names without using metadata. In this regard, we must be cautious to note that our data and results still remain quite preliminary. To improve their reliability, one should integrate other data sources and utilize them for better filtering and analysis. Semantic analysis of search results would also be of great help in this regard, though at the cost of computational complexity.

Finally, we note that our data set and results may have been influenced significantly by the particular choices we made when collecting primary data of researchers' names and research topics. We selected highly interdisciplinary conferences that were familiar to us as the source of information because of technical reasons described earlier, but we cannot eliminate the possibility of potential biases made by these choices we made. Conducting much larger-scale data collection and analysis, starting with different sets of conferences/researchers/research topics, will be necessary to reduce the effects of potential biases, which is beyond the scope of this paper.

Through this work, we aimed to illustrate a novel methodology for characterizing interdisciplinarity of researchers and their research topics by reconstructing individual-level network data about relationships among them using a general-purpose web search engine.

While our technique is still preliminary with significant room for improvement and further validation, we believe it has at least two major advantages. First, web search engine-based methods like ours make it possible for everyone to have access to a large amount of data through very simple interfaces with great flexibility. Second, the methodology can be generalized from researchers' networks to virtually any kind of networks of things based on their conceptual or cognitive similarity, as long as nodes can be represented by keywords. We have published on other applications of this method elsewhere (33,34). We believe that such web search-based research methods will become more commonly used for scholarly research in the coming years.

**Appendix**

To create lists of research topics and researchers' names, we had several student volunteers exhaustively explore the following conference websites manually and collect keywords or phrases related to scientific research as well as names of all researchers involved, including organizers, committee members, session chairs, presenters, authors/coauthors, and panelists. Each conference website was explored by at least two independent students. These conferences were chosen primarily because of our own familiarity with them (which was necessary in order to check and manually correct the raw data coming from the student volunteers).

*Mathematical Biology:*
- SMB 2006    http://www.siam.org/meetings/ls06/
- SMB 2007    http://abacus.bates.edu/~mgreer/smb_jsmb_2007/
- SMB 2008    http://www.fields.utoronto.ca/programs/scientific/CMM/08-09/SMB/index.html
- SMB 2009    http://www.math.ubc.ca/Research/MathBio/SMB2009/

*Artificial Life:*
- ALIFE X (2006)    http://www.alifex.org/
- ECAL 2007    http://www.ecal2007.org/
- ALIFE XI (2008)    http://alifexi.alife.org/
- ECAL 2009    http://www.ecal2009.org/

*Network Science:*
- NetSci 2006    http://vw.indiana.edu/netsci06/
- NetSci 2007    http://www.nd.edu/~netsci/
- NetSci 2008    http://www.ifr.ac.uk/netsci08/
- NetSci 2009    http://www.netsci09.net/

*Bio-inspired Information Technology:*
- BIONETICS 2006    http://www.bionetics.org/2006/
- BIONETICS 2007    http://www.bionetics.org/2007/
- BIONETICS 2008    http://www.bionetics.org/2008/
- BIONETICS 2009    http://www.bionetics.org/2009/

The keywords and phrases related to scientific research collected by the students were manually edited and compiled by the authors into the following list of 40 research topics used in this study. They were sorted into four major categories and the numbers of words were made similar across those categories in order to maintain a good balance among different disciplines.

[Biological and medical sciences, 13 words]
*biology*
*cancer*
*cell*
*disease*
*ecology*
*ecosystem*
*evolution*
*gene*
*immunology*
*molecule*
*neuron*
*physiology*
*protein*

[Physical sciences, 9 words]
*chemistry*
*complex network*
*complex system*
*dynamics*
*emergence*
*mathematics*
*modeling*
*physics*
*simulation*

[Engineering and robotics, 9 words]
*algorithm*
*application*
*cognition*
*computer*
*control*
*engineering*
*motor*
*robot*
*software*

[General terms, 9 words]
*communication*
*economy*
*information*
*intelligence*
*network*
*science*
*signal*
*social network*
*system*

In collecting researchers' names, the student volunteers were instructed to omit middle initials and replace any letters with diacritics (e.g., accents, umlauts, etc.) by normal alphabet letters without diacritics. These rules were implemented so as to make data inspection and correction easier. The collected lists of researchers' names were then aggregated into a single text file and carefully examined by the authors to correct any recognizable mistakes (misspellings, duplications, etc.). As a result, we had over 4,050 names collected in total.

To conduct systematic web searches using Google, we wrote a simple program code in Java that sequentially searches for each pair of keywords stored in a plain text file and then extracts and records their relatedness (i.e., number of search hits) from the search result. The code was developed using with Google Web Search API just for internal use for this study and not for public release, but it is available from the authors upon request.

The original list of researchers' names was too long for our data collection (i.e., the number of pairs, 4,050 choose 2, was beyond eight million), so we first conducted systematic web searches for each researcher's name only (together with the filtering word "research"). Then we sorted the results in terms of search hits and selected the top 1,000 researchers for the main data collection and analysis.

For the main data collection, each search query (in which "research" was always included) was searched three times at different times in a day in order to improve the reliability of search results. A known problem in using Google search results is that it sometimes returns an orders of magnitude larger (or smaller) number as search hits, which would strongly influence if the results were averaged as is. To avoid such errors, we discarded either the largest or smallest result, whichever was farther away from the median result, before calculating a mean. If both were equally distant from the median, none of them was discarded, i.e., the mean became the median.

We used programming language Python and its NetworkX module (31) for network visualization and analysis, and also Wolfram Research's Mathematica for statistical analysis. Codes for these analyses were also developed for internal use only, but can be shared upon request.

**Acknowledgments**

The authors thank the following student volunteers in the Department of Bioengineering at Binghamton University who helped the data collection task: Shawn Augustine, Tasha Casagni, Tomasz Falkowski, Joseph Perez-Rogers, Mark Schede, Josh Smiedy, Elizabeth Thompson, Han Wang, Aristo Wong, and Wendy Yung. The authors also thank Google for the data and their Web Search API which made this study possible.

**Author Contributions Statement**

HS and JA collaboratively designed the research. JA developed the data collection software. HS and JA conducted data collection and preprocessing. HS conducted detailed data analysis and wrote the manuscript text and prepared all the figures except for Fig. 3(a), which was prepared by JA. Both authors reviewed the manuscript.

**Figures and Figure Legends**

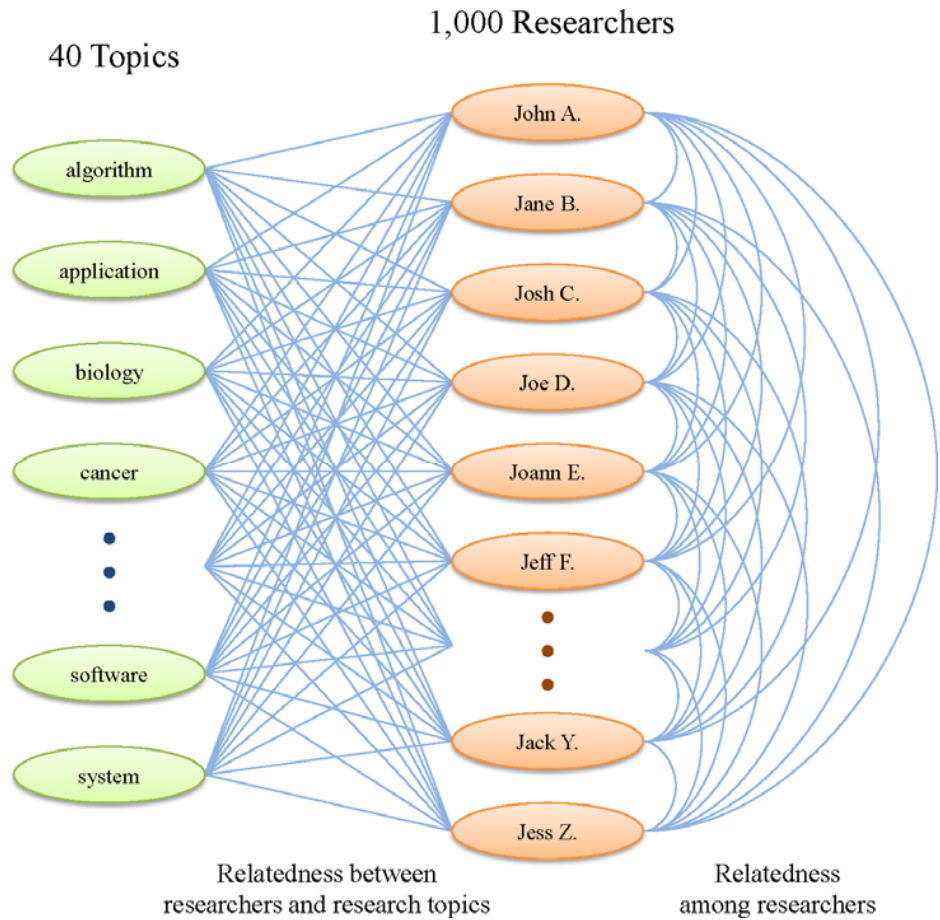

*Figure 1: An overview of the network that consists of 1,000 researchers and 40 research topics reconstructed in this study. Each link is weighted by the average number of web search hits for a search query of (research topic) + (researcher's name) + "research" (left), or (researcher's name 1) + (researcher's name 2) + "research" (right).*

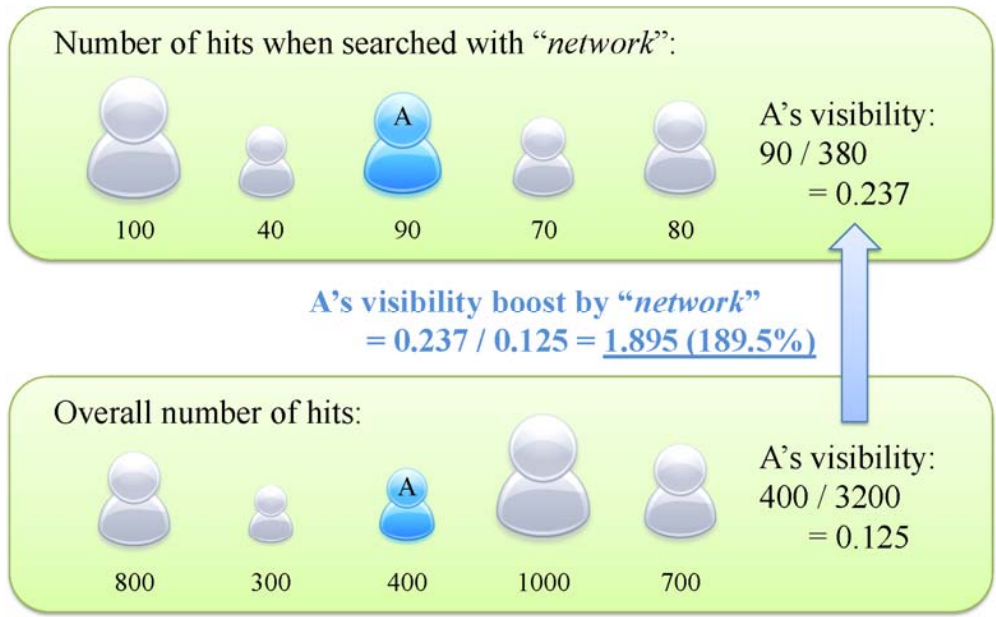

*Figure 2: An example of "visibility boost" calculation. This figure shows how to calculate the visibility boost by research topic "network" for the researcher A in the middle (blue).*

(a) 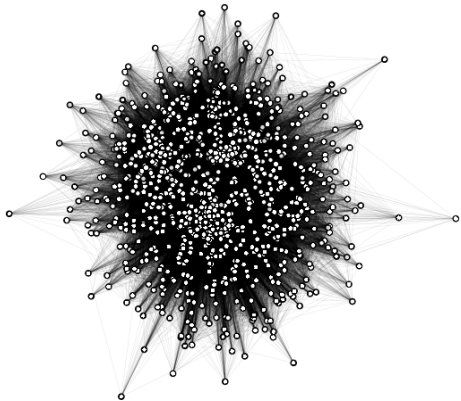 (b) 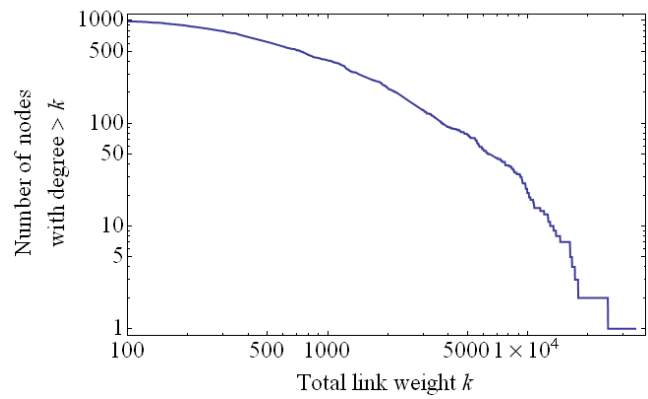

*Figure 3: (a) Reconstructed network of 1,000 researchers. (b) Distribution of total link weights of nodes, plotted as a complementary cumulative distribution function (CCDF).*

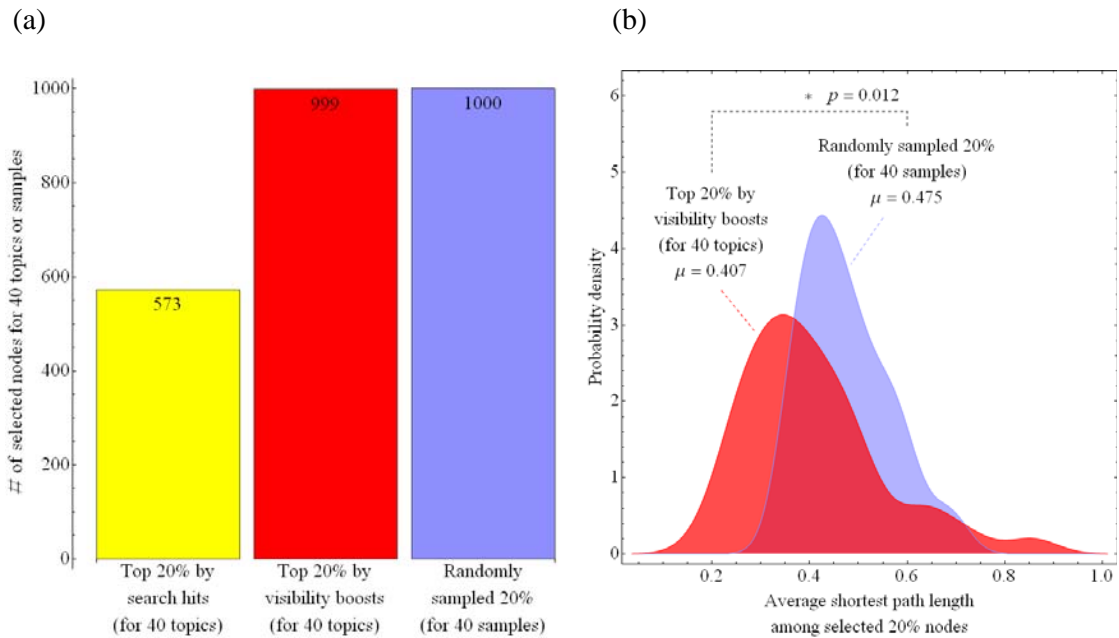

*Figure 4: (a) Comparison of the numbers of nodes (researchers) that appeared at least once in 40 times of selection trials. Yellow: Top 20% nodes selected based on their original search hit counts for each of the 40 research topics. Red: Top 20% nodes selected based on their visibility boosts by each of the 40 research topics. Blue: Random selection of 20% nodes repeated 40 times. (b) Smoothed histograms of average shortest path lengths among the selected 20% nodes in the researchers' network (N = 40 for each histogram; each sample point corresponds to one measurement of average shortest path length among the selected 20% nodes). To calculate path lengths, the reciprocals of link weights were used as edge distances. The average shortest path lengths among the top 20% nodes selected based on their visibility boosts were significantly smaller than random counterparts (p < 0.05 by standard t-test), showing that researchers who share high visibility boosts by the same topic tended to come closer to each other in the network.*

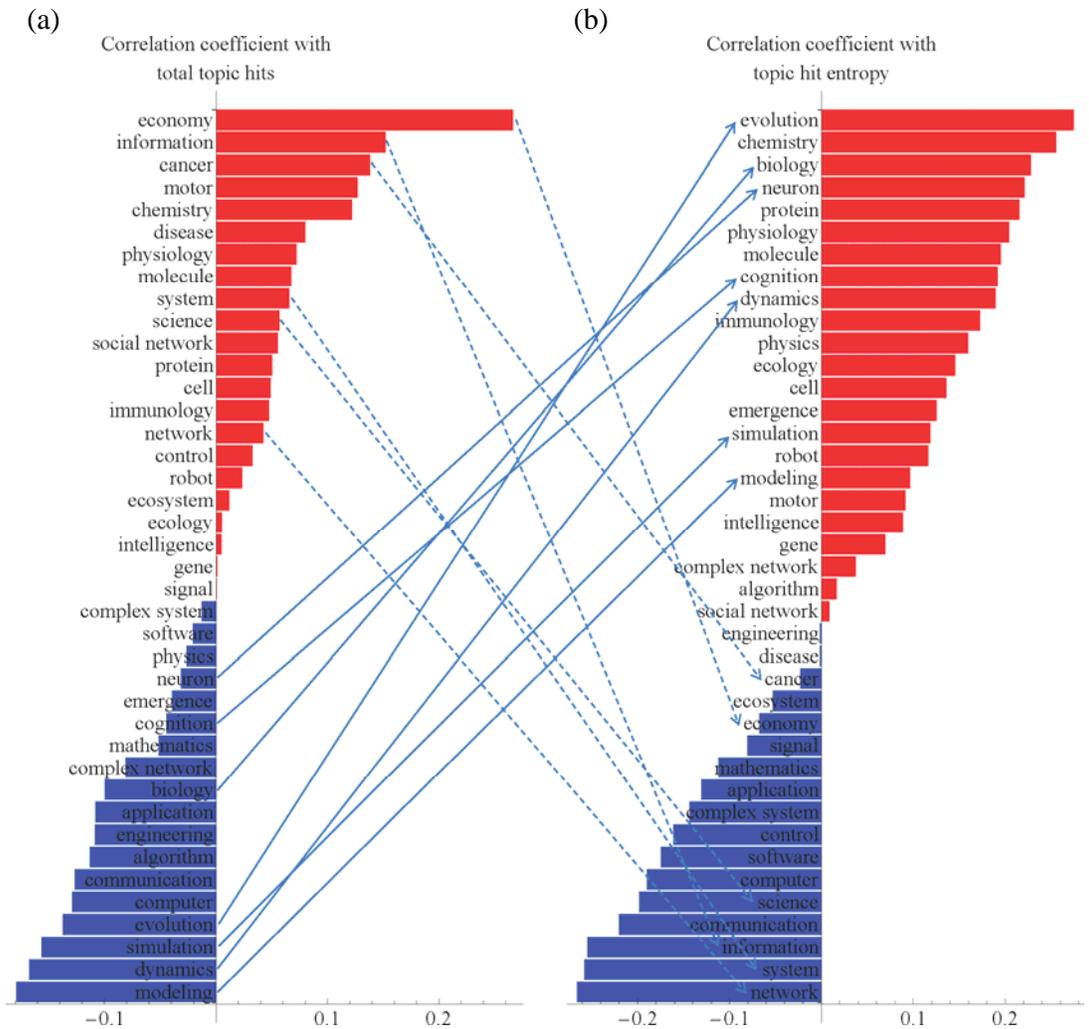

*Figure 5: Correlations between the visibility boost of each research topic and a researcher's overall popularity (total topic his, (a)) and individual-level interdisciplinarity (topic hit entropy, (b)). Upward or downward moves of topics from (a) to (b) by 20 or more places in the ranking are indicated by solid and dashed arrows, respectively.*

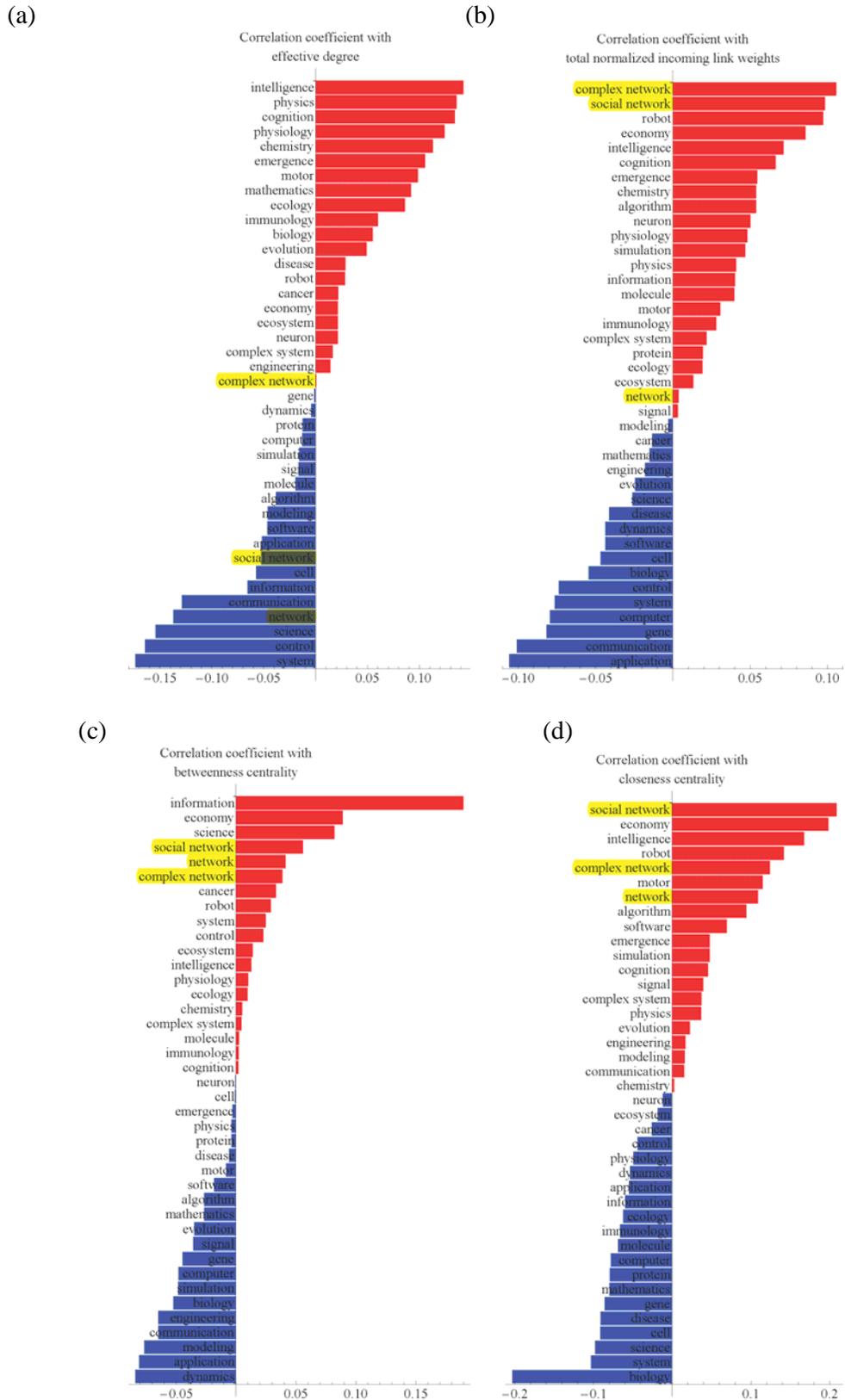

*Figure 6: Correlation between each research topic and a researcher's social-level interdisciplinarity. (a) Effective degree. (b) Total normalized incoming link weights. (c) Betweenness centrality. (d) Closeness centrality. Network-related topics are highlighted.*

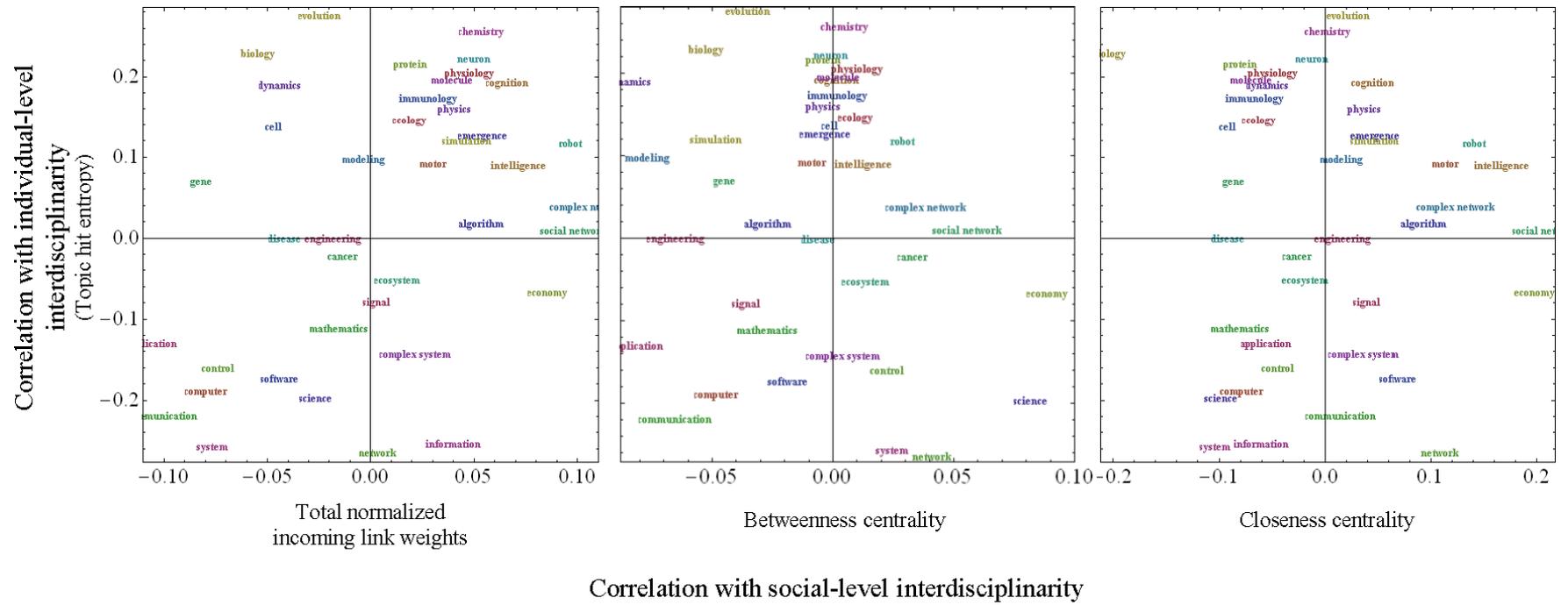

*Figure 7: Two dimensional maps summarizing each research topic's correlations with a researcher's individual-level interdisciplinarity (vertical, topic hit entropy) and social-level interdisciplinarity (horizontal, three centrality measurements).*